\documentclass[sigconf]{acmart}
\usepackage{graphicx}

\usepackage{xcolor}
\usepackage{amsfonts}
\usepackage{amsmath}
\usepackage{booktabs}
\usepackage{colortbl}
\usepackage{multirow}
\usepackage{rotating}
\usepackage{todonotes}

\newcommand{\greyrule}{\arrayrulecolor{black!30}\midrule\arrayrulecolor{black}}
\newcommand\sys{\textsc{GRIT}}
\newcommand\qs{\textsc{TQE}}
\AtBeginDocument{%
  }

\copyrightyear{2025}
\acmYear{2025}
\setcopyright{acmlicensed}\acmConference[WWW Companion '25]{Companion Proceedings of the ACM Web Conference 2025}{April 28-May 2, 2025}{Sydney, NSW, Australia}
\acmBooktitle{Companion Proceedings of the ACM Web Conference 2025 (WWW Companion '25), April 28-May 2, 2025, Sydney, NSW, Australia}
\acmDOI{10.1145/3701716.3717859}
\acmISBN{979-8-4007-1331-6/2025/04}

\begin{document}

\title{GRIT: Graph-based Recall Improvement \\ for Task-oriented E-commerce Queries}
  

\author{Hrishikesh Kulkarni}
\affiliation{%
  \institution{Information Retrieval Lab, Georgetown University}
  \city{Washington}
  \state{DC}
  \country{USA}}
\email{first@ir.cs.georgetown.edu}

\author{Surya Kallumadi}
\affiliation{%
  \institution{coursera.org}
  \city{Atlanta}
  \state{Georgia}
  \country{USA}}
\email{first@ksu.edu}

\author{Sean MacAvaney}
\affiliation{%
  \institution{University of Glasgow}
  \city{Glasgow}
  \country{United Kingdom}}
\email{first.last@glasgow.ac.uk}

\author{Nazli Goharian}
\affiliation{%
  \institution{Information Retrieval Lab, Georgetown University}
  \city{Washington}
  \state{DC}
  \country{USA}}
\email{first@ir.cs.georgetown.edu}

\author{Ophir Frieder}
\affiliation{%
  \institution{Information Retrieval Lab, Georgetown University}
  \city{Washington}
  \state{DC}
  \country{USA}}
\email{first@ir.cs.georgetown.edu}

\renewcommand{\shortauthors}{Kulkarni et al.}

\begin{abstract}
Many e-commerce search pipelines have four stages, namely: retrieval, filtering, ranking, and personalized-reranking. The retrieval stage must be efficient and yield high recall because relevant products missed in the first stage cannot be considered in later stages. This is challenging for task-oriented queries (queries with actionable intent) where user requirements are contextually intensive and difficult to understand.
To foster research in the domain of e-commerce, we created a novel benchmark for Task-oriented Queries (\qs{}) by using LLM, which operates over the existing ESCI product search dataset. Furthermore, we propose a novel method `Graph-based Recall Improvement for Task-oriented queries' (\sys{}) to address the most crucial first-stage recall improvement needs. \sys{} leads to robust and statistically significant improvements over state-of-the-art lexical, dense, and learned-sparse baselines. Our system supports both traditional and task-oriented e-commerce queries, yielding up to $6.3\%$ recall improvement.
In the indexing stage, \sys{} first builds a product-product similarity graph using user clicks or manual annotation data. During retrieval, it locates neighbors with higher contextual and action relevance and prioritizes them over the less relevant candidates from the initial retrieval. This leads to a more comprehensive and relevant first-stage result set that improves overall system recall. 
Overall, \sys{} leverages the locality relationships and contextual insights provided by the graph using neighboring nodes to enrich the first-stage retrieval results. 
We show that the method is not only robust across all introduced parameters, but also works effectively on top of a variety of first-stage retrieval methods. 
\end{abstract}



\maketitle

\section{Introduction}
To manage efficiency-effectiveness trade-offs, Information Retrieval (IR) systems often employ a two-stage ranking pipeline: an inexpensive (but imprecise) first stage followed by a precise (but most costly) second stage reranker~\cite{matveeva2006high}. The first stage is often a lexical Bag-of-Words (BOW) model~\cite{Robertson2009ThePR}, a learned-sparse model~\cite{10.1145/3477495.3531857}, or a dense bi-encoder~\cite{karpukhin-etal-2020-dense}. Meanwhile, the re-ranker is often a neural pre-trained cross-encoder that can refine the initial results~\cite{10.1145/3331184.3331340}. Two-stage pipelines yield impressive results when used to solve standard IR problems; unfortunately, the e-commerce domain adds additional complications, namely the need for detailed understanding of specifics and the tacit applicability of potentially available products
\cite{10.1145/3642979.3643007,10.1145/3626772.3661373,10.1145/3397271.3401446}. Hence, additional stages are introduced in e-commerce product search due to business-specific promotions and personalization
\cite{10.1145/3451964.3451966,10.1145/3539618.3591859}. The product knowledge and  product relationships are crucial in e-commerce product retrieval and hence knowledge about customer activity data plays a key role in the whole process \cite{10.1145/3336191.3371778}. 
Categorically, domain-specific vocabulary, the scale of available user-product interaction data, and efficiency requirements due to multistage pipeline differentiate e-commerce product search from general purpose search \cite{10.1145/3451964.3451966,10.1145/3642979.3643007}. 

Importantly, there are task-oriented queries (e.g., `Locate a replacement AC compressor for 2006 Honda Odyssey', `Find a suitable Apple fast charger compatible with my devices', etc.) where actionable intent is at the center of the query. The actionable intent focused on a particular task makes these queries contextually intensive. These queries are receiving increasing attention due to their specificity and action-centric nature \cite{hantask,qin-etal-2023-end,10.1145/3269206.3269326}.
IR suffers when handling task-oriented queries in e-commerce due to their action orientation and domain-specific contextually intensive nature. Complicating matters, there is no standard benchmark or dataset for task-oriented queries for algorithm development and evaluation \cite{yim2024taskorientedqueriesbenchmarktoqb}. To address this issue, we present a Task-oriented Query set for e-commerce (\qs{}) benchmark on the widely used ESCI Product Search data \cite{reddy2022shopping} as an evaluation forum for IR approaches in e-commerce product search.

\begin{figure*}
\centering



\includegraphics[scale=0.45]{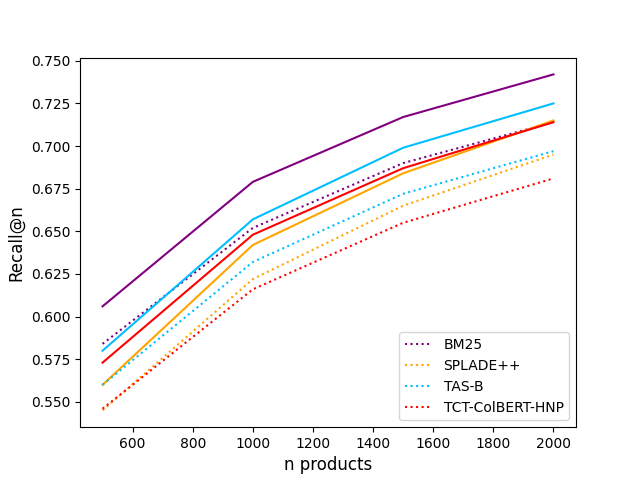}
\includegraphics[scale=0.45]{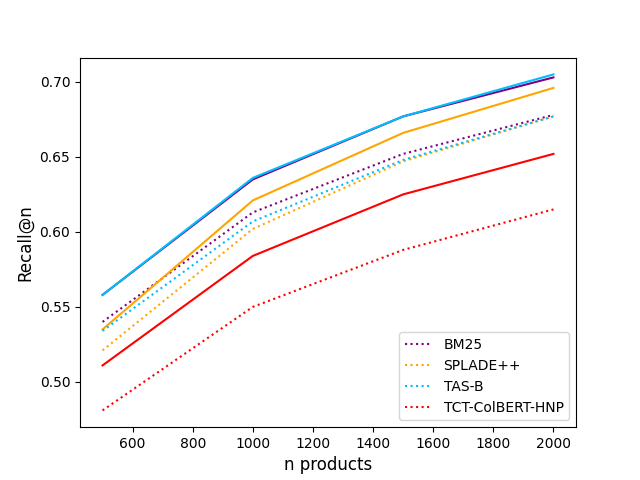}

\caption{Comparison of our approach (\sys{}) with baselines in terms of Recall at $n$ results. Solid line: \sys{}, Dotted line: baseline for respective baseline. The plots are on traditional ESCI queries and \qs{} benchmark (small) respectively. It is clearly evident that \sys{} significantly improves Recall of initial retrieval across methods, $n$ values and query types.
}
\label{fig:teasor}
\end{figure*}

In e-commerce product search, the first stage (retrieval stage) is responsible for retrieving a set of relevant products from the product catalog using minimal computational resources. The later stages use sophisticated machine learning (ML) algorithms and filtering mechanisms to determine their exact positions in the ranking order \cite{10.1145/3539618.3591859}. 
Vast attention has focused on later stages of the pipeline, particularly on personalizing results and providing country/region-specific rankings to yield better results \cite{lu-etal-2021-graph}.  As subsequent stages are built on top of the first stage, missing relevant products in the first stage is prohibitive.
We focus on that particular recall improvement by introducing GRIT, an approach
which uses the unexplored locality and neighborhood knowledge derived from
a product-product similarity graph. Figure \ref{fig:teasor} compares the performance of \sys{} with the baselines across number of top search results considered. 

Our contributions are threefold: 
\begin{itemize}
 \item We present a Task-oriented Query set for e-commerce (\qs{}) benchmark on the widely used ESCI Product Search data as an evaluation forum for IR approaches in e-commerce product search.
\item We introduce \sys{} which builds a product-product similarity graph using user-click/train-annotation data to identify neighbors with higher relevance. 
 \item We present experiments that demonstrate that \sys{} significantly improves recall for both task-oriented and traditional e-commerce queries.
\end{itemize}

\section{Related Work}

IR search systems aim to execute user-provided queries to retrieve documents that are relevant to them.
The process is challenging when the information need expressed by the query is unclear or indirect 
\cite{pmlr-v119-guu20a,10.1145/3394486.3403305,10.1016/j.ipm.2021.102522,wen2024elaborativesubtopicqueryreformulation}. 

\subsection{Lexical and Neural First-stage IR}

To retrieve the most relevant information or candidates, traditional text-based information retrieval systems rely on matching terms between the query and the documents \cite{Croft2009SearchEI}. These methods use an inverted index, which leads to very high efficiency. They were further improved in terms of effectiveness by considering various term weighting schemes. This advancement led to a variety of TF-IDF models exemplified by BM25 \cite{Robertson2009ThePR}. These bag-of-words based approaches like BM25 gave impetus to early efforts focused on efficiency aspects of the retrieval problem. These research initiatives continued with multiple variations of BM25 to tackle different application specific challenges \cite{Robertson2009ThePR}. 

Neural models work on a semantic level, overcoming lexical overlap limitations of traditional methods. With improvement in computational resources while maintaining low-latency requirements of the e-commerce domain, single-representation bi-encoder models have become an alternative for the first-stage retrieval. TAS-B \cite{10.1145/3404835.3462891}, TCT-ColBERT-HNP \cite{lin-etal-2021-batch}, RetroMAE \cite{RetroMAE} and SimLM \cite{wang-etal-2023-simlm} are some of the state-of-the-art single-representation bi-encoder models.
Learned-sparse models exemplified by SPLADE++ \cite{10.1145/3477495.3531857,10.1145/3404835.3463098} use regularization to learn sparse representations of documents and queries. This enables efficiency comparable to lexical methods using inverted index while maintaining high effectiveness of neural methods.
Three types of methods (lexical, bi-encoder neural, and learned-sparse) are popularly used for first-stage retrieval in e-commerce.

\subsection{IR for E-Commerce and Task-oriented Queries}

E-commerce connects customers to different products and services. E-commerce product search deals with finding and ranking relevant products for user queries. The search for e-commerce products needs to be handled differently due to the prevalence of domain-specific vocabulary, seasonality, and the scale of the available user-product interaction data \cite{10.1145/3642979.3643007}. Information retrieval in e-commerce is more challenging due to the multiple heterogeneous sources of information, the wide spectrum of intent of users, and vocabulary gaps \cite{10.1145/3209978.3210185}.  Hence, various kinds of signals, including clicks, and favorites, need to be integrated into the model to derive relevance \cite{10.1145/3209978.3209993}. Furthermore, e-commerce demands high efficiency due to multi-stage pipeline with additional business specific promotional initiative layers \cite{10.1145/3451964.3451966}. Prior research has shown that traditional retrieval systems generally outperform general purpose pretrained embedding models \cite{campos2023overviewtrec2023product}. Attempts are reported where mismatched results were used for contextual retrieval improvements \cite{nguyen-etal-2020-learning}. Furthermore, it is necessary to leverage procedural knowledge which is crucial for responding to task-oriented e-commerce queries \cite{10.1145/2766462.2767744}. 

\begin{figure*}
    \centering
\includegraphics[width=0.7\linewidth]{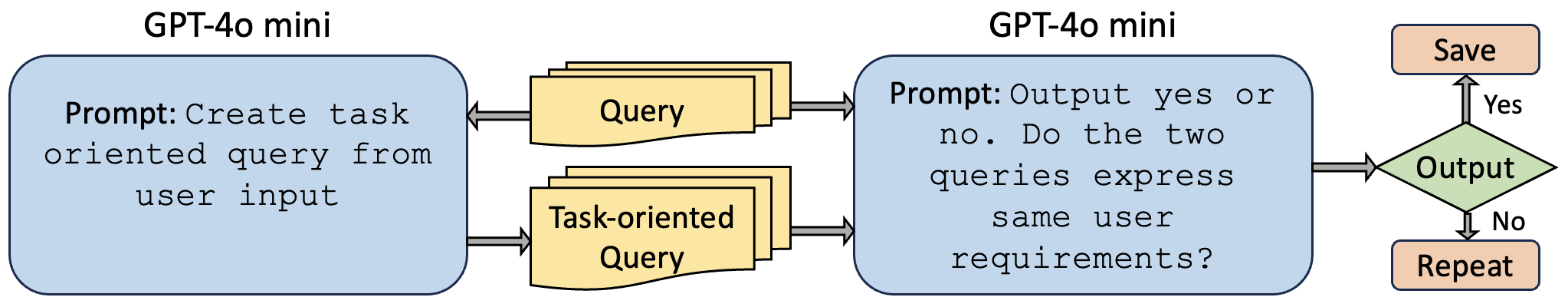}
    \caption{\qs{}: Task-oriented query generation}
    \label{fig:tqe}
\end{figure*}

Task-oriented queries or action driven queries enhance the e-commerce product search process \cite{10.1145/3583780.3615474} and are also important in the web search \cite{6040862}. Hence, it is crucial to assess task-oriented queries \cite{yim2024taskorientedqueriesbenchmarktoqb}. Existing benchmarks in relevant NLP fields focus on task-oriented dialogue systems \cite{raghu-etal-2021-unsupervised,yim2024taskorientedqueriesbenchmarktoqb}. The previous research though underlines some efforts to create task-oriented queries in a generalized setup - there is a need to have a specific task-oriented queries benchmark dataset focused on e-commerce. 
   
\subsection{Neighborhood-based Search}
A variety of systems leverage a document-level nearest-neighbor structure, referred to as a \textit{corpus graph}. For instance, HNSW \cite{8594636} is the most popular Approximate Nearest Neighbor (ANN) search method, which traverses a multi-level graph to retrieve results for vector-based queries. In addition, methods such as GAR \cite{macavaney:cikm2022-adaptive} and LADR \cite{10.1145/3539618.3591715} perform ANN search by leveraging a corpus graph and an inexpensive first stage to identify candidate documents. Unlike HNSW, GAR and LADR can approximate any relevance model, but are ultimately upper-bounded by the model's effectiveness. Meanwhile, methods such as LexBoost \cite{10.1145/3685650.3685658} focus on improving lexical retrieval with graph-based score formulations.
In addition, weighted bipartite query-document graphs have been used to include user query characteristics in graph navigation \cite{frayling:ecir2024-reverted}.

In contrast with these existing methods that use lexical or semantic signals to construct a corpus graph, we propose \sys{}, which uses user click/train annotation data to build a product-product similarity graph. Furthermore, \sys{} is not upper bounded by the recall of first-stage retrieval and works on all types of first-stage retrieval methods. In addition, it does not increase the candidate size, thus having no effect on the latency of consecutive stages.

\section{Benchmark: \qs}

We create and introduce the first Task-oriented Query set for e-commerce (\qs{}) benchmark consisting of 22472 English queries. This novel benchmark is based on the ESCI data for improving product search \cite{reddy2022shopping}.
TQE differs from existing task-oriented benchmarks as it specifically focuses on e-commerce retrieval unlike TOQB \cite{yim2024taskorientedqueriesbenchmarktoqb} which focuses on dialog systems.
We first identify English queries with US locale from the ESCI product search data. As shown in Figure \ref{fig:tqe}, we then use GPT-4o mini \cite{gpt4omini} to generate equivalent task-oriented queries from the ESCI product search English query set. This is performed by prompting GPT-4o mini to create a task-oriented query from the input traditional query. We choose GPT-4o mini LLM as a result of its high effectiveness at relatively lower costs \cite{gpt4omini}. In addition, we have a validation layer where we validate the consistency of expressed user requirements in the traditional and generated task-oriented query. The generation process is repeated until the task-oriented query clears the validation. Validation is performed by prompting GPT-4o mini to output yes or no based on whether the generated task-oriented query and the original query express the same user requirements. Both the query generation and validation prompts (see Fig. \ref{fig:tqe}) are decided after manually trying out a wide range of alternatives. Action words like find, locate, add etc are present in the final query stressing on the task-oriented nature. As the user requirements in each ESCI product search query and the corresponding generated task-oriented query are the same, we can use gold labels (qrels) from the ESCI data to evaluate the results on the \qs{} benchmark.

The original ESCI product search data contains around 130 thousand unique queries and 2.6 million manually labeled (query, product) relevance judgments. The dataset is multilingual with queries in English, Spanish, and Japanese.  We filter the queries and focus only on the English subset. The test set with US locale has a small query set with 8954 English queries and a large query set with 22472 English queries.
We release both small and large task-oriented query sets to evaluate state-of-the-art approaches for e-commerce setting.

\section{Method: \sys{}}

\begin{figure*}
    \centering
    \includegraphics[width=0.7\linewidth]{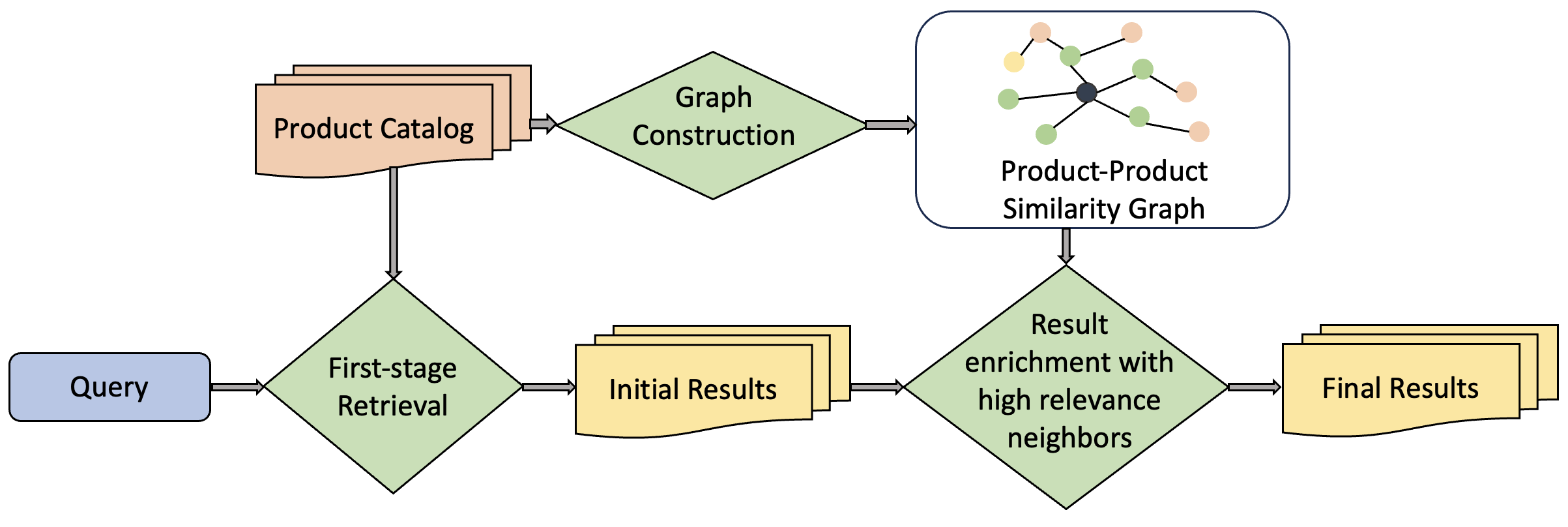}
    \caption{\sys{} System Architecture}
    \label{fig:sys-arch}
\end{figure*}

\sys{} consists of two parts: building a product-product similarity graph at the index creation stage and graph-based retrieval as shown in Figure \ref{fig:sys-arch}. As the graph construction is performed at the index creation stage and neighbors are determined before retrieval, this approach is highly scalable. Hence, the improvements shown by \sys{} are at virtually no additional latency overhead.

\subsection{Graph building approach:}

The train set of ESCI product search data \cite{reddy2022shopping} has products judged on four point scale namely: 
\begin{itemize}
    \item Exact (E): The product is exactly relevant to the query.
    \item Substitute (S): The product is somewhat relevant to the query.
    \item Complement (C): The product could be used in combination with a relevant product to the query.
    \item Irrelevant (I): The product is irrelevant to the query.
\end{itemize}

We use `Exact', `Substitute' and `Complement' labeled products for each query in the training set to create a corpus graph.
For each query in the training set, we iterate over every pair of products in the annotations ($A$) leading to $9$ possible combinations shown in Table \ref{tab:graph}.
This is based on the Cluster Hypothesis derivation that the products relevant to a query are similar to each other \cite{Jardine1971TheUO}.
Considering every pair of products in this set to be neighbors, we calculate edge weight based on the weight matrix shown in Table \ref{tab:graph} and the number of times this combination of products repeats in the train set query annotations. 

The weight of an edge $w_{pq}$ between products $p$ and $q$ can be determined using the weight matrix $wm$ and labels ($l$) of $p$ and $q$ in the annotations. The final weight in the graph is the sum of the weights assigned over all $(p, q)$ pair occurrences.
Based on this, $w_{pq}$ is formally defined in equation \ref{eq:ce}. 

\begin{equation}
    w_{pq} = \sum_{(p,q)\in A}wm[l_p,l_q]
    \label{eq:ce}
\end{equation}


The weight matrix values (Table \ref{tab:graph}) are chosen based on decreasing relevance of `Exact' then `Substitute' then `Complement' labels. For e.g. E-E, E-S and E-C are 3, 2, 1 respectively. Further, we set C-C to 1 as if product1 and product2 complement the requirements of the query then it does not mean that they are exact neighbors of each other. For e.g. `running headband' and `track pants' products are both complementary to the query `running shoes'. `Running headband' and `track pants', though relevant to each other, are not exact neighbors, and hence the weight 1 is assigned. Based on similar intuition, other weights in the matrix are determined. This is similar to how reward values are determined in Reinforcement Learning. Though there are exceptions for the above logic (for e.g. Exact products for ambiguous queries might not be relevant to each other), the final weight value ($w_{pq}$) won't be high as their co-occurrence is rare. Here, we note that the final edge weight between products p and q is $w_{pq}$ which is a function of both the weight matrix $wm$ and the number of co-occurrences of p and q across the dataset. Here, we do not consider products labeled Irrelevant (I) as they do not provide any similarity insights with respect to other products.

\begin{table}[]
\setlength{\tabcolsep}{10pt}
    \centering
    \begin{tabular}{l c c c}
    \toprule
       & E & S & C \\
    \midrule
     E & 3 & 2 & 1 \\
     S & 2 & 2 & 1 \\
     C & 1 & 1 & 1 \\ 
    \bottomrule\\
    \end{tabular}
    \caption{Weight matrix $wm$ for product-product pairs. Here E, S, C are from the four point scale from ESCI product search data viz. E is Exact, S is Substitute, C is Complement and I is Irrelevant. $wm[E,S]$ denotes the weight for the pair with annotations E and S. }
    \label{tab:graph}
\end{table}

\begin{table*}
\centering
\caption{Examples of from \qs{} benchmark with corresponding ESCI product search query.}
\small
\begin{tabular}{p{0.3\linewidth} | p{0.55\linewidth}}
\toprule
Traditional ESCI Query & Generated Task-oriented Query \\
\midrule
06 honda odyssey a/c compressor	& Locate a replacement AC compressor for a 2006 Honda Odyssey \\
\greyrule
08 civic si fog lights without switch & Find a wiring diagram or installation guide for 2008 Civic Si fog lights that do not have switch \\
\greyrule
1 bushel laundry basket &	Find a single bushel laundry basket for purchase online \\
\greyrule
100\% success without college &	Research strategies for achieving success without a college degree\\
\greyrule
10mls amber serum vials without flip caps & Find suppliers for 10ml amber serum vials that do not have flip caps\\
\greyrule
apple fast charger & Find a suitable Apple fast charger that is compatible with my devices\\
\bottomrule
\end{tabular}
\label{tab:toq}
\end{table*}

\subsection{Retrieval Approach}
\sys{} first retrieves $n$ results per query with the given baseline. Here, $n$ is the number of products expected to be in the first-stage retrieval output. We then use the top $t\%$ results from the initial retrieval as seed nodes in the graph. $t$ is our seed size parameter.
The higher quality of seeds leads to more relevant neighbors, making $t$ an important parameter to analyze.
We rank the neighbors based on their final weights, which are the sum of their respective weights for every occurrence. 
The weight of a neighbor $q$ ($w_q$) is the sum of the weights of all the edges $q$ have with the products in the result set $R$.
Hence, $w_q$ can be formally defined by equation \ref{eq:pq}. 

\begin{equation}
    w_{q} = \sum_{(p)\in R}w_{pq}
    \label{eq:pq}
\end{equation}


We use the new shortlisted neighbors equivalent to the bottom $b\%$ of the initial retrieval and replace the products of lower relevance. Here, $b$ is our product replacement parameter. In order to ensure that more relevant neighbors replace less relevant lower ranked products, tuning the $b$ parameter is crucial. 

\section{Experiment}
In this work, we address the following research questions:
\begin{description}
  \item[RQ1:] Does user-click/train-annotation data based graph provide necessary quality insights regarding product-product similarity?
  \item[RQ2:] Can LLMs simulate task-oriented queries effectively for e-commerce?
  \item[RQ3:] How does the performance of retrieval methods compare using traditional ESCI queries and equivalent generated task-oriented queries?
  \item[RQ4:] Does \sys{} improve Recall of state-of-the-art first-stage retrieval methods for traditional and task-oriented e-commerce queries?
  \item[RQ5:] How does a change in the seed size parameter $t$ and product replacement parameter $b$ affect the performance of \sys{}?
\end{description}

For reproducibility, our code and task-oriented query set are available here.\footnote{https://github.com/Georgetown-IR-Lab/TQE-GRIT} 

\subsection{Models and Parameters}

\sys{} uses user-click/train-annotation data to build the product-product similarity graph. With the `ESCI product search data', we have a significant train query set with 68139 English queries with relevance judgements (E, S, C, I). We use product pairs with relevance judgements E, S, and C in a weighted fashion to determine neighbors and edge-weights for the product-product similarity graph. At industry level, these data could easily be replaced with in-session user-click data. Product-product similarity graph construction is a one-time process executed at indexing stage.

We try `seed size parameter $t$' at $0.01$, $0.02$, $0.03$ and $0.04$ for extensive evaluation. Here, $t$ determines the percentage of the top results from the initial retrieval to be used as seeds in the graph. We also evaluate it across a wide range of `product replacement parameters $b$' including $0.1$, $0.2$, $0.3$ and $0.4$. Here, $b$ determines the percentage of bottom results to be replace with neighbors. We also evaluate \sys{} across various types of first-stage retrieval baselines for traditional and task-oriented queries.

\subsection{Baselines and Implementation}
To present a comprehensive study, we evaluated \sys{} on the state-of-the-art lexical method BM25, the neural methods TAS-B and TCT-ColBERT-HNP, and the learned sparse method SPLADE++. BM25 and SPLADE++ use an inverted index and hence are highly efficient. This makes them a popular choice for first-stage retrieval in e-commerce. However, improved computational resources make the use of single representation biencoder methods such as TAS-B and TCT-ColBERT-HNP for first-stage retrieval possible, making them compelling baseline choices.   
In addition, seed nodes are chosen from the most relevant first-stage retrieval products to identify neighbors of greater relevance. These identified neighbors are used to replace products from the initial retrieval with lower relevance while maintaining the candidate size. Hence, comparisons with baselines are ideal. \sys{} is implemented on the baselines listed above for a wide range of parameter values. Our focus is on evaluating the Recall at different candidate sizes.

\section{Results and Analysis}

\begin{table}
\centering 
\caption{Recall Improvement on ESCI query set (small) by \sys{} for different b values. In method with \sys{} - method is the baseline initial retrieval. $b$ is the product replacement parameter. $^\dagger$ denotes statistical significance (paired t-test: $p<0.05$). \textbf{Bold} denotes best performance. Statistically significant and robust improvements in Recall are evident across retrieval methods and product replacement parameter values. For example in TCT-Col.-HNP with \sys{} and $b=0.3$, the Recall@1000 performance of TCT-ColBERT-HNP improves from $0.616$ to $0.648$ leading to $5.2\%$ improvement.}
\small
\begin{tabular}{llrrrr}
\toprule
Method & b & R@500 & R@1000 & R@1500 & R@2000 \\
\midrule
\midrule
BM25 & -- & 0.584 & 0.652 & 0.690 & 0.714 \\
\greyrule
\multirow{ 3}{*}{\shortstack[l]{BM25 \\with \sys{}}} & 0.1 & 0.599$^\dagger$ & 0.671$^\dagger$ & 0.709$^\dagger$ & 0.735$^\dagger$ \\
 & 0.2 & 0.605$^\dagger$ & 0.677$^\dagger$ & 0.716$^\dagger$ & 0.741$^\dagger$ \\
 & 0.3 & \textbf{0.606}$^\dagger$ & \textbf{0.679}$^\dagger$ & \textbf{0.717}$^\dagger$ & \textbf{0.742}$^\dagger$ \\
\greyrule
Improvement & -- & 3.8\% & 4.1\% & 3.9\% & 3.9\% \\
\midrule
\midrule
SPLADE++ & -- & 0.545 & 0.622 & 0.665 & 0.695 \\
\greyrule
\multirow{ 3}{*}{\shortstack[l]{SPLADE++ \\with \sys{}}} & 0.1 & 0.556$^\dagger$ & 0.636$^\dagger$ & 0.679$^\dagger$ & 0.710$^\dagger$ \\
 & 0.2 & 0.559$^\dagger$ & 0.641$^\dagger$ & 0.684$^\dagger$ & 0.714$^\dagger$ \\
 & 0.3 & \textbf{0.560}$^\dagger$ & \textbf{0.642}$^\dagger$ & \textbf{0.684}$^\dagger$ & \textbf{0.715}$^\dagger$ \\
\greyrule
Improvement & -- & 2.8\% & 3.2\% & 2.9\% & 2.9\% \\
\midrule
\midrule
TAS-B & -- & 0.560 & 0.632 & 0.672 & 0.697 \\
\greyrule
\multirow{ 3}{*}{\shortstack[l]{TAS-B \\with \sys{}}} & 0.1 & 0.574$^\dagger$ & 0.650$^\dagger$ & 0.691$^\dagger$ & 0.718$^\dagger$ \\
 & 0.2 & 0.579$^\dagger$ & 0.656$^\dagger$ & 0.698$^\dagger$ & 0.724$^\dagger$ \\
 & 0.3 & \textbf{0.580}$^\dagger$ & \textbf{0.657}$^\dagger$ & \textbf{0.699}$^\dagger$ & \textbf{0.725}$^\dagger$ \\
\greyrule
Improvement & -- & 3.6\% & 4.0\% & 4.0\% & 4.0\% \\
\midrule
\midrule
TCT-Col.-HNP & -- & 0.546 & 0.616 & 0.655 & 0.681 \\
\greyrule
\multirow{ 3}{*}{\shortstack[l]{TCT-Col.-HNP \\with \sys{}}} & 0.1 & 0.564$^\dagger$ & 0.638$^\dagger$ & 0.677$^\dagger$ & 0.705$^\dagger$ \\
 & 0.2 & 0.571$^\dagger$ & 0.645$^\dagger$ & 0.685$^\dagger$ & 0.713$^\dagger$ \\
 & 0.3 & \textbf{0.573}$^\dagger$ & \textbf{0.648}$^\dagger$ & \textbf{0.687}$^\dagger$ & \textbf{0.714}$^\dagger$ \\
\greyrule
Improvement & -- & 4.9\% & 5.2\% & 4.9\% & 4.8\% \\
\bottomrule
\end{tabular}
\label{tab:s1}
\end{table}

\begin{table}
\centering
\caption{Recall Improvements on \qs{} benchmark (small) by \sys{} for different $b$ values. In method with \sys{} - method is the baseline initial retrieval. $b$ is the product replacement parameter. $^\dagger$ denotes statistical significance (paired t-test: $p<0.05$). \textbf{Bold} denotes best performance. Statistically significant and robust improvements in Recall are evident across retrieval methods and product replacement parameter values. For example in TCT-Col.-HNP with \sys{} and $b = 0.3$, the Recall@1500 performance of TCT-ColBERT-HNP improves from $0.588$ to $0.625$ leading to $6.3$\% improvement.}
\small
\begin{tabular}{llrrrr}
\toprule
Method & b & R@500 & R@1000 & R@1500 & R@2000 \\
\midrule
\midrule
BM25 & -- & 0.540 & 0.613 & 0.652 & 0.678 \\
\greyrule
\multirow{ 3}{*}{\shortstack[l]{BM25  \\with \sys{}}} & 0.1 & 0.554$^\dagger$ & 0.629$^\dagger$ & 0.670$^\dagger$ & 0.696$^\dagger$ \\
 & 0.2 & 0.557$^\dagger$ & 0.634$^\dagger$ & 0.675$^\dagger$ & 0.702$^\dagger$ \\
 & 0.3 & \textbf{0.558}$^\dagger$ & \textbf{0.635}$^\dagger$ & \textbf{0.677}$^\dagger$ & \textbf{0.703}$^\dagger$ \\
\greyrule
Improvement & -- & 3.3\% & 3.6\% & 3.8\% & 3.7\% \\
\midrule
\midrule
SPLADE++ & -- & 0.521 & 0.602 & 0.647 & 0.677 \\
\greyrule
\multirow{ 3}{*}{\shortstack[l]{SPLADE++ \\with \sys{}}} & 0.1 & 0.532$^\dagger$ & 0.616$^\dagger$ & 0.661$^\dagger$ & 0.692$^\dagger$ \\
 & 0.2 & 0.535$^\dagger$ & 0.620$^\dagger$ & 0.665$^\dagger$ & 0.695$^\dagger$ \\
 & 0.3 & \textbf{0.535}$^\dagger$ & \textbf{0.621}$^\dagger$ & \textbf{0.666}$^\dagger$ & \textbf{0.696}$^\dagger$ \\
\greyrule
Improvement & -- & 2.7\% & 3.2\% & 2.9\% & 2.8\% \\
\midrule
\midrule
TAS-B & -- & 0.534 & 0.607 & 0.648 & 0.677 \\
\greyrule
\multirow{ 3}{*}{\shortstack[l]{TAS-B \\with \sys{}}} & 0.1 & 0.551$^\dagger$ & 0.627$^\dagger$ & 0.668$^\dagger$ & 0.697$^\dagger$ \\
 & 0.2 & 0.556$^\dagger$ & 0.634$^\dagger$ & 0.675$^\dagger$ & 0.704$^\dagger$ \\
 & 0.3 & \textbf{0.558}$^\dagger$ & \textbf{0.636}$^\dagger$ & \textbf{0.677}$^\dagger$ & \textbf{0.705}$^\dagger$ \\
\greyrule
Improvement & -- & 4.5\% & 4.8\% & 4.5\% & 4.1\% \\
\midrule
\midrule
TCT-Col.-HNP & & 0.481 & 0.550 & 0.588 & 0.615 \\
\greyrule
\multirow{ 3}{*}{\shortstack[l]{TCT-Col.-HNP \\with \sys{}}} & 0.1 & 0.500$^\dagger$ & 0.573$^\dagger$ & 0.613$^\dagger$ & 0.640$^\dagger$ \\
 & 0.2 & 0.508$^\dagger$ & 0.582$^\dagger$ & 0.623$^\dagger$ & 0.650$^\dagger$ \\
 & 0.3 & \textbf{0.511}$^\dagger$ & \textbf{0.584}$^\dagger$ & \textbf{0.625}$^\dagger$ & \textbf{0.652}$^\dagger$ \\
\greyrule
Improvement & -- & 6.2\% & 6.2\% & 6.3\% & 6.0\% \\
\bottomrule
\end{tabular}
\label{tab:s2}
\end{table}
\begin{table*}
\centering
\caption{Recall Improvement on large query sets by \sys{}. \sys{} is applied on the respective baseline initial retrieval method. $^\dagger$ denotes statistical significance (paired t-test: $p<0.05$). \textbf{Bold} denotes best performance. Statistically significant and robust improvements in Recall are evident across retrieval methods and product replacement parameter values. For example in TAS-B, \sys{} improves Recall@1000 from $0.708$ to $0.734$ leading to $3.7\%$ improvement for Traditional ESCI queries. Similarly, for Task-oriented queries, in TAS-B, \sys{} improves Recall@1000 from $0.678$ to $0.706$ leading to $4.1\%$ improvement.}
\setlength{\tabcolsep}{6pt}
\small
\begin{tabular}{lrrrrrrrr}
\toprule
Method & R@500 & Imp. & R@1000 & Imp. & R@1500 & Imp. & R@2000 & Imp. \\
\hline\\
\multicolumn{9}{c}{Traditional ESCI Product Search Queries (Large Set)}\\
\hline
BM25 & 0.658 & & 0.721 & & 0.754 & & 0.775 & \\
\sys{} & \textbf{0.682}$^\dagger$ & 3.6\% & \textbf{0.748}$^\dagger$ & 3.7\% & \textbf{0.782}$^\dagger$ & 3.7\% & \textbf{0.803}$^\dagger$ & 3.6\%\\
\greyrule
SPLADE++ & 0.622 & & 0.695 & & 0.734 & & 0.760 & \\
\sys{} & \textbf{0.637}$^\dagger$ & 2.4\% & \textbf{0.714}$^\dagger$ & 2.7\% & \textbf{0.752}$^\dagger$ & 2.5\% & \textbf{0.778}$^\dagger$ & 2.4\% \\
\greyrule
TAS-B & 0.642 & & 0.708 & & 0.743 & & 0.766 & \\
\sys{} & \textbf{0.663}$^\dagger$ & 3.3\% & \textbf{0.734}$^\dagger$ & 3.7\% & \textbf{0.770}$^\dagger$ & 3.6\% & \textbf{0.793}$^\dagger$ & 3.5\% \\
\greyrule
TCT-ColBERT-HNP & 0.629 & & 0.694 & & 0.729 & & 0.753 & \\
\sys{} & \textbf{0.657}$^\dagger$ & 4.5\% & \textbf{0.726}$^\dagger$ & 4.6\% & \textbf{0.761}$^\dagger$ & 4.4\% & \textbf{0.784}$^\dagger$ & 4.1\% \\
\hline\\
\multicolumn{9}{c}{\qs{} benchmark (Large Set)}\\
\hline
BM25 & 0.608 & & 0.676 & & 0.713 & & 0.736 & \\
\sys{} & \textbf{0.627}$^\dagger$ & 3.1\% & \textbf{0.700}$^\dagger$ & 3.6\% & \textbf{0.737}$^\dagger$ & 3.4\% & \textbf{0.761}$^\dagger$ & 3.4\% \\
\greyrule
SPLADE++ & 0.590 & & 0.667 & & 0.709 & & 0.737 & \\
\sys{} & \textbf{0.604}$^\dagger$ & 2.4\% & \textbf{0.685}$^\dagger$ & 2.7\% & \textbf{0.727}$^\dagger$ & 2.5\% & \textbf{0.755}$^\dagger$ & 2.4\% \\
\greyrule
TAS-B & 0.608 & & 0.678 & & 0.715 & & 0.740 & \\
\sys{} & \textbf{0.633}$^\dagger$ & 4.1\% & \textbf{0.706}$^\dagger$ & 4.1\% & \textbf{0.744}$^\dagger$ & 4.1\% & \textbf{0.769}$^\dagger$ & 3.9\% \\
\greyrule
TCT-ColBERT-HNP & 0.554 & & 0.622 & & 0.658 & & 0.682 & \\
\sys{} & \textbf{0.587}$^\dagger$ & 6.0\% & \textbf{0.659}$^\dagger$ & 5.9\% & \textbf{0.696}$^\dagger$ & 5.8\% & \textbf{0.721}$^\dagger$ & 5.7\% \\

\bottomrule
\end{tabular}
\label{tab:l}
\end{table*}

We now discuss our results and how they address the 
RQs.

\subsection{RQ1: Quality of user-click/train-annotation data based graph}
RQ1 is designed to understand the importance of click-data in providing insights regarding product-product similarity. To verify it, the recall improvements are analyzed with and without \sys{}. As evident in Table \ref{tab:s1}, \ref{tab:s2} and \ref{tab:l}, a higher number of relevant results are obtained after replacing the initial results of lower relevance with neighbors of products with higher relevance. This behavior is attributed to the clustering hypothesis, and we conclude that neighbors of a product in the graph are indeed similar to the product. 
As evident in Table \ref{tab:l}, for large query set, TCT-ColBERT-HNP Recall@1000 results improve from 0.694 to 0.726 (traditional queries) while they improve from 0.622 to 0.659 on the TQE benchmark. Similar trends are observed across first-stage retrieval methods, datasets, and evaluation metrics, demonstrating the utility of the graph.
Hence, product-product similarity graph built on the training set annotation data is of high quality.

\subsection{RQ2: LLMs as task-oriented query generators}

RQ2 is designed to verify the credibility of the approach of using LLM to generate task-oriented  queries. The approach has some support from the literature, where LLMs are used to generate queries for zero-shot settings \cite{1004646} and generalized task-oriented queries \cite{yim2024taskorientedqueriesbenchmarktoqb}.
We generate task-oriented queries using GPT-4o mini \cite{gpt4omini} on the traditional ESCI product search query set. Along with LLM-based validation layer, an annotator has manually validated the quality of these generated queries on $100$ randomly sampled task-oriented and traditional query pairs. The criteria for validation was expressing exact same user requirements and having a task-oriented nature. In $98$ out of $100$ cases, the generated queries express the exact same user requirements, while all $100$ queries have a task-oriented nature. Some sample queries can be found in Table \ref{tab:toq}. For example, the queries `06 honda odyssey a/c compressor' (traditional) and `Locate a replacement AC compressor for a 2006 Honda Odyssey' (LLM generated) have the exact same requirements but the latter one has task-oriented nature. 
Thus addressing RQ2.

\subsection{RQ3: The comparison of SOTA methods on traditional vs. task-oriented queries}
RQ3 is designed to understand the query type-specific performance of state-of-the-art methods when we have traditional and task-oriented queries in focus. The purpose is to understand the performance implications and compromises, if any, in such a case. As evident in Tables \ref{tab:s1} and \ref{tab:s2}, there is on average $5.9\%$ drop in the recall effectiveness of state-of-the-art methods on task-oriented queries when compared with traditional queries for small query set. Furthermore, as evident in Table \ref{tab:l}, the drop is on average $6.2\%$ for a large query set. 
The trend of drop in Recall effectiveness when evaluated on task-oriented queries can be observed across all four baselines. We also note that among the baselines, BM25 performs the best followed by TAS-B. This can be attributed to domain-specific vocabulary.
As evident in Table \ref{tab:s1}, Recall@1000 for BM25 on traditional small query set is $0.652$. Under the same setting, as per Table \ref{tab:s2}, Recall@1000 is $0.613$ for the equivalent task-oriented small query set. Hence, a drop of $0.039$ can be observed. Similarly, for the same setting, Recall@1000 for TAS-B drops from 0.632 to 0.607. Similar trends have been observed on a large query set with BM25 Recall@1000 dropping from $0.721$ to $0.676$ and TAS-B from $0.708$ to $0.678$ under the same setting.
The generated task-oriented queries express the exact user requirements as the corresponding ESCI product search queries. The loss in effectiveness can be attributed to the task-oriented nature of the generated queries, which makes it a bit more difficult, even for the state-of-the-art methods to understand the user requirements addressing RQ3.

\subsection{RQ4: The recall effectiveness of \sys{} across query types}
\begin{figure*}
\centering



\includegraphics[scale=0.47]{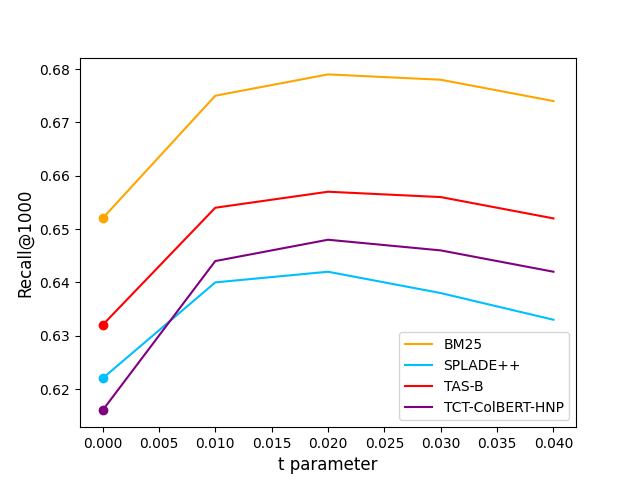}
\includegraphics[scale=0.47]{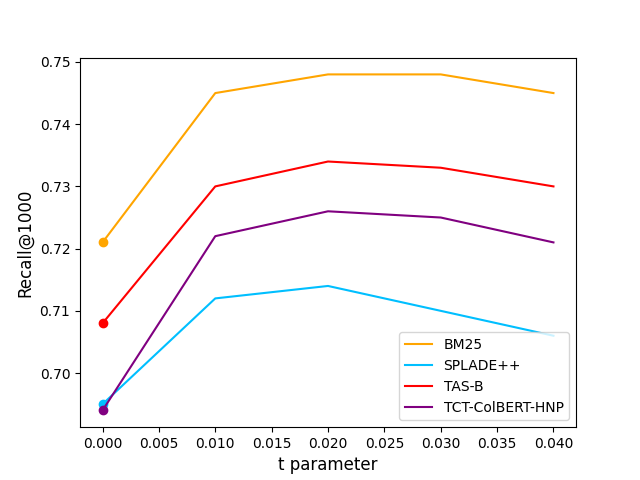}
\\(a.) On traditional small query set. \hspace{96pt} (b.) On traditional large query set.

\caption{Recall at first 1000 results for \sys{} $t$ parameter values from $t=0.000$ to $t=0.040$ at the interval of $0.005$. Baseline result is at $t=0$ denoted by dot. On ESCI small and large query sets respectively. Robustness of \sys{} is clearly evident across seed size parameter ($t$) values on traditional ESCI product search queries. Further, we also note that BM25 performs the best followed by TAS-B.}
\label{fig:tgraphs1}
\end{figure*}

RQ4 is designed to understand the extent of the recall improvements \sys{} brings to the table in the case of traditional and task-oriented queries. \sys{} shows statistically significant improvements over both small and large sets of traditional ESCI product search queries (see Table \ref{tab:s1} and \ref{tab:l}). These improvements are robust across parameters, which is also evident in Figure \ref{fig:tgraphs1}.
For the small query set, Recall@1000 results for BM25 improve from $0.652$ to $0.679$, while for TAS-B they improve from $0.632$ to $0.657$. 
Similar trends are observed on the large query set for traditional queries.
Even in the case of task-oriented queries, it shows statistically significant and robust improvements (see Table \ref{tab:s2} and \ref{tab:l}). 
For the small query set, Recall@1000 results for BM25 improve from $0.613$ to $0.635$, while for TAS-B they improve from $0.607$ to $0.636$. Similar trends are observed on the large query set of \qs{}.
This validates that \sys{} is effective at improving recall across query types addressing RQ4.
Overall, \sys{} leads to an average improvement of $3.9\%$ and $4.2\%$ on traditional and task-oriented queries, respectively.

\begin{figure*}
\centering



    
\includegraphics[scale=0.47]{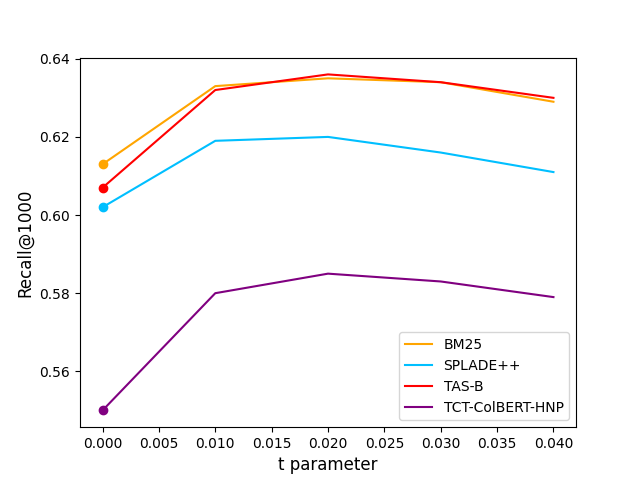}
\includegraphics[scale=0.47]{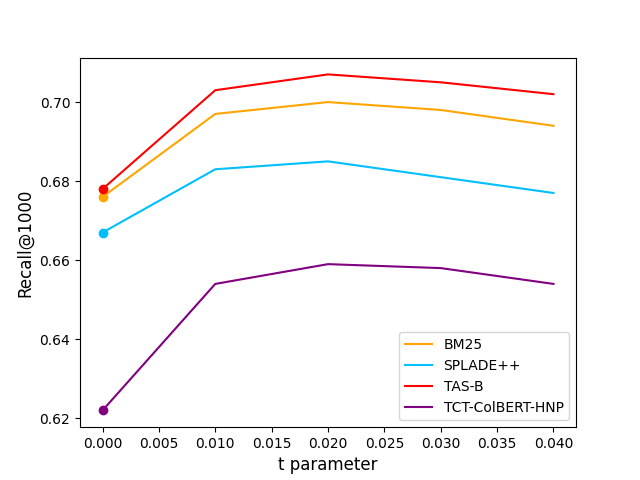}
\\(a.) On task-oriented small query set. \hspace{96pt} (b.) On task-oriented large query set.

\caption{Recall at first 1000 results for \sys{} across $t$ parameter values from $t=0.000$ to $t=0.040$ at the interval of $0.005$. Baseline result is at $t=0$ denoted by dot. On \qs{} benchmark - small and large query sets respectively. Robustness of \sys{} is clearly evident across seed size parameter ($t$) values on the \qs{} benchmark. Further, we also note that for the small query set BM25 and TAS-B show comparable performance. But, in case of large query set TAS-B outperforms BM25.}
\label{fig:tgraphs2}
\end{figure*}

\subsection{RQ5: Robustness across variations in $t$ and $b$ parameters}

RQ5 is designed to understand the nature of the impact and sensitivity of seed size and product replacement parameters on the performance of \sys{}. As evident in Table \ref{tab:s1}, \ref{tab:s2} and \ref{tab:l}, the results across $b$ values are statistically significant. With increase in the $b$ parameter value we observe increase in the Recall effectiveness. For traditional queries (small set) for $b$ parameter values in $0.1$, $0.2$, and $0.3$, Recall@1000 improves from $0.652$ to $0.671$, $0.677$ and $0.679$ respectively for BM25 with \sys{}. Similarly for TAS-B with \sys{}, Recall@1000 improves from $0.632$ to $0.650$, $0.656$ and $0.657$ with increasing $b$ parameter values respectively. We observe similar trends for task-oriented queries (small set). As $b$ parameter value is increased to $0.1$, $0.2$, and $0.3$, BM25 with \sys{}, Recall@1000 effectiveness improves from $0.613$ to $0.629$, $0.634$ and $0.635$ respectively. While for TAS-B with \sys{}, it improves from $0.607$ to $0.627$, $0.634$ and $0.636$ respectively. Furthermore, we observe that the results saturate after $b$ value reaches $0.3$. These observations are evident across query sets and query types. 

Similarly, as evident in Figure \ref{fig:tgraphs1} and \ref{fig:tgraphs2}, the results are also robust across $t$ values.
Here, we note that $t$ parameter value of $0.02$ leads to the best Recall@1000 effectiveness across all four baselines and both query sets. 
For the tuned parameter values of $b = 0.3$ and $t = 0.02$ we observed best Recall results with BM25, SPLADE++, TAS-B and TCT-ColBERT-HNP leading to Recall@1000 of 0.679, 0.642, 0.657 and 0.648 for traditional queries (small set). Similarly, they lead to Recall@1000 of 0.635, 0.621, 0.636 and 0.584 for task-oriented queries (small set) respectively. For large sets of both types of queries, the trends are similar with best results and improvements shown in Table \ref{tab:l}.
From this we can infer that top $2\%$ form the right seed size which lead to high quality neighbors addressing RQ5.

\section{Conclusion and Future Work}
In this paper, we presented a Task-oriented Query set for e-commerce (\qs{}) benchmark using LLM as an evaluation forum for IR approaches in e-commerce product search. \qs{} will enable the further algorithm development and evaluation on the most required task-oriented e-commerce setup. We also proposed a novel approach \sys{} that effectively uses the associations among products to refine the candidates retrieved at the first stage by performing an optimal and selective replacement. It embeds user activity data into node relationships and exploits rich locality-based knowledge to provide recall boost to first stage retrieval.  Importantly, it provides statistically significant and robust recall improvements when used on top of different state-of-the-art retrieval approaches. \sys{} works effectively on both traditional and task-oriented queries. 

As a future work, \sys{} can be extended to improve results in case of applications where multiple products are to be purchased for a given task.

\bibliographystyle{ACM-Reference-Format}
\bibliography{sample-new}

\end{document}